\documentclass[aps,amsmath,amssymb, reprint, prl, preprintnumbers, twocolumn]{revtex4-2}
\pdfoutput=1

\usepackage{graphicx}
\usepackage{epstopdf}
\usepackage{amsmath, amssymb}

\usepackage{hyperref}
\usepackage{bbm,array,amsfonts,graphicx,wrapfig,float,mathtools,multirow}
\usepackage[dvipsnames]{xcolor}

\usepackage{tikz, float}
\usetikzlibrary{patterns,shapes.misc}

\newcommand{\be}{\begin{equation}}
\newcommand{\ee}{\end{equation}}
\newcommand{\beq}{\begin{equation}}
\newcommand{\beql}[1]{\begin{equation}\label{#1}}
\newcommand{\eeq}{\end{equation}}
\newcommand{\ba}{\begin{array}}
\newcommand{\ea}{\end{array}}
\newcommand{\bea}{\begin{eqnarray}}
\newcommand{\beal}[1]{\begin{eqnarray}\label{#1}}
\newcommand{\eea}{\end{eqnarray}}
\newcommand{\ben}{\begin{enumerate}}
\newcommand{\een}{\end{enumerate}}
\newcommand{\bean}{\begin{eqnarray*}}
\newcommand{\eean}{\end{eqnarray*}}
\newcommand{\eref}[1]{(\ref{#1})}

\newcommand{\nn}{\nonumber}

\newcommand{\fref}[1]{Figure \ref{#1}}

\def\d{\delta}

\def\BZ{\mathbb{Z}}

\begin{document}

\title{Integrable Systems for Generalized Toric Polygons and Higgsed $5d$ $\mathcal{N}=1$ Theories}

\preprint{UNIST-MTH-26-RS-03} 
\preprint{CGP26005}

\author{Minsung Kho}
\email[\texttt{minsung@unist.ac.kr}]{}
\affiliation{
Department of Mathematical Sciences, Ulsan National Institute of Science and Technology,
50 UNIST-gil, Ulsan 44919, South Korea
}

\author{Kimyeong Lee}
\email[\texttt{klee@bimsa.cn}]{}
\affiliation{
Beijing Institute of Mathematical Sciences and Applications (BIMSA), Huaibei Town, Huairou
District, Beijing 101408, China
}

\author{Norton Lee}
\email[\texttt{norton.lee@ibs.re.kr}]{}
\affiliation{
Center for Geometry and Physics, Institute for Basic Science (IBS),
Pohang 37673, South Korea
}

\author{Rak-Kyeong Seong}
\email[\texttt{seong@unist.ac.kr}]{}
\affiliation{
Department of Mathematical Sciences, and Department of Physics, Ulsan National Institute of Science and Technology,
50 UNIST-gil, Ulsan 44919, South Korea
}

\begin{abstract}
The interplay between toric Calabi-Yau $3$-folds, dimer integrable systems, and $5$-dimensional quantum field theories has proved fruitful. 
We extend this framework to generalized toric polygons (GTPs) and show that their integrable systems arise from refined birational transformations of known dimer integrable systems acting on the Casimirs and Hamiltonians as well as the Poisson structure and spectral curves. 
We argue that these transformations are realized as Hanany-Witten transitions producing $(p,q)$ $5$-brane webs dual to GTPs. 
We show that the resulting $5d$ $\mathcal{N}=1$ theory is obtained by Higgsing a higher-rank theory whose associated toric Calabi-Yau has a toric diagram of the same shape as the GTP.
\end{abstract} 

\maketitle

%========================================================
\section{Introduction}

The study of integrable systems
has led to a broad range of insights into
$4d$ $\mathcal{N}=2$ supersymmetric gauge theories \cite{Seiberg:1994aj}.
An important feature of $4d$ $\mathcal{N}=2$ theories is that
they admit an uplift to $5d$ $\mathcal{N}=1$ theories \cite{Seiberg:1996bd, Morrison:1996xf, Intriligator:1997pq, Closset:2019juk},
which can be realized as Type IIB 5-brane webs \cite{Aharony:1997bh, Aharony:1997ju, Leung:1997tw}.
When such a $5d$ $\mathcal{N}=1$ theory is geometrically engineered in M-theory
via a toric Calabi-Yau 3-fold \cite{fulton1993introduction},
the 5-brane webs -- also referred to as $(p,q)$-web diagrams -- are dual to the Newton polygon
of the associated toric Calabi-Yau 3-fold.
In this way, the story comes full circle:
the toric Calabi-Yau 3-fold underlying the $(p,q)$-web also 
corresponds to a cluster integrable system \cite{2011arXiv1107.5588G, Eager:2011dp, Franco:2015rnr, Hatsuda:2015qzx, Bershtein:2017swf, Marshakov:2019vnz, Huang:2020neq, Lee:2023wbf, Lee:2024bqg, Bershtein:2024lvd}.

The correspondence between cluster integrable systems and toric Calabi-Yau 3-folds
stems from the fact that the worldvolume theory on a D3-brane probing such a toric Calabi-Yau
is a $4d$ $\mathcal{N}=1$ supersymmetric gauge theory \cite{Douglas:1997de, Douglas:1996sw, Feng:2000mi, Feng:2001xr}.
Under T-duality, the probe D3-branes map to D5-branes suspended from an NS5-brane that wraps a holomorphic curve $\Sigma$,
which yields a Type IIB configuration known as a brane tiling \cite{Franco:2005rj, Hanany:2005ve, Franco:2005sm}.
A brane tiling consists of a bipartite periodic graph on a 2-torus and is also
referred to as a dimer model \cite{2003math.....10326K}.
Directed paths along the edges of this bipartite graph generate a path algebra whose generators
provide canonical coordinates for an integrable system,
which we refer to here as a dimer integrable system.

It was recently argued in \cite{Kho:2025fmp, Kho:2025jxk} that
if two toric Calabi-Yau 3-folds with reflexive polygons \cite{1993alg.geom.10003B, Hanany:2012hi} as their toric diagrams are related by a birational transformation,
then the associated dimer integrable systems are birationally equivalent \cite{kollar1998birational, 2012arXiv1212.1785A, Ghim:2024asj, Ghim:2025zhs}.
A birational transformation $\varphi_A$
acts on the Newton polynomial $P(x,y)$ of a toric diagram $\Delta$.
Its zero locus,
$\Sigma: P(x,y)=0$ defines the holomorphic curve, also known as the Calabi-Yau mirror curve \cite{Hori:2000ck, Feng:2005gw}, 
wrapped by the NS5-brane in the Type IIB brane configuration represented by a brane tiling.
Under a suitable identification of coefficients, the same curve $\Sigma$
is the spectral curve of the dimer integrable system,
which is written in terms of Hamiltonians associated with internal points of $\Delta$
together with Casimirs associated with external points.

%------------------------------------------------------------------------------------------
\begin{figure*}[htt!!]
\centering
\includegraphics[width=0.95\linewidth]{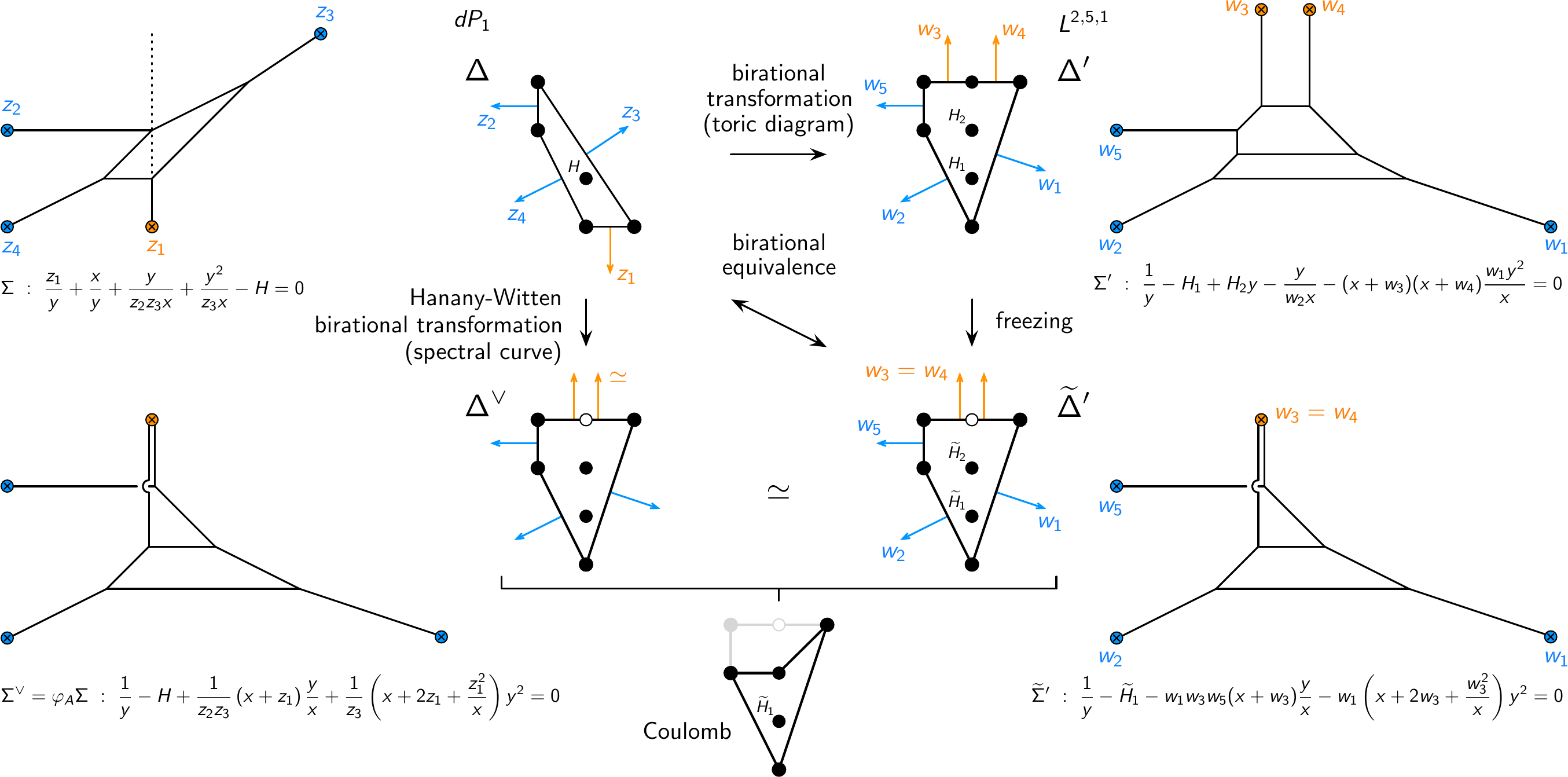}
\caption{
The toric diagrams $\Delta$ of the $dP_1$ model and $\Delta^\prime$ of the $L^{2,5,1}$ model are related by a birational transformation.
Under a Hanany-Witten transition, the $(p,q)$-web dual to $\Delta$ becomes
a $(p,q)$-web where two external $5$-branes end on the $7$-brane, making the dual polygon $\Delta^\vee$ a GTP.
Because $\Delta^\vee$ is of the same shape as $\Delta^\prime$, 
we can identify the two $5$-branes terminating on the same $7$-brane with zig-zag paths $w_3$ and $w_4$ in the brane tiling of $L^{2,5,1}$.
In the corresponding dimer integrable system, we can impose $w_3 = w_4$ resulting in a reduced (frozen) integrable system that describes the dynamics of the $5d$ $\mathcal{N}=1$ theory corresponding to the GTP $\Delta^\vee$. 
We also note that the reduced integrable system is also birationally equivalent to the dimer integrable system corresponding to $\Delta$.
}
\label{fig_01}
\end{figure*}
%------------------------------------------------------------------------------------------

The birational equivalence is realized by a refined birational map
that matches Casimirs and Hamiltonians, identifies the spectral curve, and preserves the Poisson structure
of cluster variables.
Accordingly, birational equivalence between dimer integrable systems 
was first introduced in \cite{Kho:2025fmp} for toric diagrams with the same number of internal points,
so that the same number of commuting Hamiltonians can be identified on both sides.
In this letter, we addresses the complementary situation:
what is the correct notion of birational equivalence between integrable systems 
when a birational transformation changes the number of internal points of the toric diagram?

Our work is motivated by the dual $(p,q)$-web description, 
which exists for every toric diagram of a toric Calabi-Yau 3-fold.
Starting from a $(p,q)$-web in which each external $5$-brane ends on a separate $7$-brane,
a Hanany-Witten transition \cite{Hanany:1996ie} generically produces brane configurations in which several external $5$-brane legs terminate on the same $7$-brane.
The object dual to such a web is no longer an ordinary toric diagram.
Instead, it is described by a \textit{generalized toric polygon} (GTP) \cite{Benini:2009gi, vanBeest:2020kou, Bourget:2023wlb, CarrenoBolla:2024fxy, Franco:2023flw, Arias-Tamargo:2024fjt, Franco:2023mkw, CarrenoBolla:2024fxy},
whose boundary edges contain white vertices indicating external $5$-brane legs ending on common $7$-branes.
In the Newton polynomial and associated spectral curve this leads to natural combinations of boundary coefficients, 
while in the integrable system it leads to a controlled reduction, or freezing, of degrees of freedom.
We show that birational maps relating ordinary toric diagrams with different numbers of internal points are naturally interpreted as
birational equivalences between an ordinary dimer integrable system and a \emph{reduced} integrable system associated with a GTP of the same shape.

In what follows we briefly recall the construction of dimer integrable systems in the form needed here,
define the reduction mechanism for a integrable system associated with a GTP from the $(p,q)$-web viewpoint,
and demonstrate the mechanism explicitly for a birational equivalence between corresponding integrable systems associated to the $dP_1$ \cite{Feng:2002zw, Feng:2001xr, Benvenuti:2006qr, Hanany:2012hi} and $L^{2,5,1}$ \cite{Cvetic:2005ft, Franco:2005sm, Benvenuti:2005ja, Butti:2005sw} geometries.

%========================================================
\section{Dimer Integrable Systems and Birational Transformations}
\label{sec:01}

A bipartite periodic graph on $T^2$, also referred to as a dimer model \cite{2003math.....10326K} or brane tiling \cite{Franco:2005rj, Hanany:2005ve, Franco:2005sm}
determines, upon choosing a fundamental domain in $T^2$,
a Kasteleyn matrix $K(x,y)$ \cite{kasteleyngraph}
whose entries are weighted sums of edges $e_{ij}$ in the brane tiling with monomials in
$(x,y) \in (\mathbb{C}^*)^2$ that encode winding numbers $w(e_{ij}) \in \mathbb{Z}^2$ of edges around $T^2$.
We use here the determinant of the Kasteleyn matrix $K(x,y)$ in order to obtain the Newton polynomial of the form,
\beal{es01a01}
P(x,y) = \mathop{\sum_{p_{(n_x,n_y)}}}_{(n_x,n_y) \in \Delta} \overline{p}_{(n_x,n_y)}~ x^{n_x} y^{n_y} ~,~
\eea
where the sum runs over perfect matchings $p_{(n_x,n_y)}$ of the brane tiling, which are
associated to vertices with coordinates $(n_x, n_y) \in \mathbb{Z}^2$ in the toric diagram $\Delta$.
Each monomial coefficient $\overline{p}_{(n_x,n_y)}$ in $P(x,y)$ is a product of $e_{ij}$ that form a perfect matching $p_{(n_x,n_y)}$ in the brane tiling.
The spectral curve of the associated dimer integrable system is the zero locus,
$\Sigma: P(x,y)=0$.

A convenient way to organize coefficients of $P(x,y)$ is to factor out a monomial coefficient $\overline{p}_0$
associated with a reference perfect matching $p_0$ \cite{Kho:2025fmp, Kho:2025jxk}
in the brane tiling, 
so that each monomial coefficient in the reduced polynomial 
$P_0(x,y) = \overline{p}_0^{-1} \cdot P(x,y)$ 
becomes a loop variable in the dimer integrable system.
These loop variables are expressed as ratios
of directed edge variables $e_{ij}^\pm$ \cite{Kho:2025fmp, Kho:2025jxk} corresponding to products of directed paths in the brane tiling.
Terms in $P_0(x,y)$ corresponding to boundary lattice points of $\Delta$ 
have monomial coefficients that are Casimirs $\delta_{(n_x,n_y)} = (\overline{p}_0)^{-1} \cdot \overline{p}_{(n_x,n_y)}$,
while terms corresponding to internal points have coefficients that are 
Hamiltonians of the form $H = \sum_u \gamma_u$, where $\gamma_u$ are 1-loops.

Let $f_i$ be face variables, also known as cluster variables, 
associated with faces of the brane tiling, 
or equivalently nodes of the corresponding quiver.
The face variables are expressed in terms of cyclic products of 
directed edge variables $e_{ij}^\pm$ 
that form a directed path counterclockwise along the boundary edges of a face. 
Within the path algebra defined by 
$(e_{ij}^\pm)^{-1} = e_{ij}^{\mp}$ and $e_{ij}^+ \cdot e_{ij}^- = 1$, 
the face variables $f_i$ satisfy the Poisson brackets
$\{f_i,f_j\} = I_{i,j} ~f_i f_j$, 
where $I_{i,j}$ is the adjacency matrix of the quiver.
As for the face variables $f_i$, 
one can also identify a finite set of directed zig-zag paths $z_r$ \cite{2003math.....10326K, 2009arXiv0901.4662B, Bocklandt:2011vt, Hanany:2015tgh, Hanany:2012vc}
that are likewise expressed as cyclic products of $e_{ij}^\pm$.
Casimirs and the 1-loops in the Hamiltonians
can be expressed as products of
$f_i$ and $z_r$.

We can expand the Newton polynomial for $\Delta$ as follows, 
\beal{es01a10}
P(x,y) = \sum_{m=-a}^{b} C_{m}(x) y^m ~,~
\eea
where $a,b>0$,
and $C_m(y)$ are sub-polynomials of $P(x,y)$ 
for $-a \leq m \leq b$.
Using this expansion, we can introduce a birational transformation of the form,
\beal{es01a11}
\varphi_A ~:~ (x,y) \mapsto (x,A(x)y) ~,~
\eea
where the Laurent polynomial $A(x)$ is chosen such that $A(x)^{-m}$ is a polynomial divisor of $C_m(x)$ for all $m$. 
Calling the resulting Newton polynomial $P^\prime(x,y)$ and the corresponding toric diagram $\Delta^\prime$, 
it was shown in \cite{Kho:2025fmp, Kho:2025jxk} that if $\Delta$ and $\Delta^\prime$ have the same number of internal vertices, 
then the corresponding dimer integrable systems are birationally equivalent.
The subtlety addressed in this work 
is that $\varphi_A$ may change the number of internal points of the toric diagram.
With this in mind, we explain why the appropriate target of the birational transformation in this case is 
a \textit{reduced} integrable system associated with a generalized toric polygon of the same shape as $\Delta^\prime$.

%========================================================
\section{Generalized toric polygons and $(p,q)$-webs}
\label{sec:02}

In a $(p,q)$-web dual to an ordinary toric diagram, 
each of the external 5-brane legs ends on a distinct 7-brane.
These external 5-brane legs are also in one-to-one correspondence with zig-zag paths of the associated brane tiling, 
where the $\mathbb{Z}^2$ direction of the external legs matches 
the winding number of the corresponding zig-zag path on $T^2$.
Under a Hanany-Witten transition \cite{Hanany:1996ie}, 
multiple external $5$-branes can end up terminating on the same $7$-brane.
The dual polygon is then no longer a toric diagram in the conventional sense 
and is instead a generalized toric polygon (GTP) \cite{Benini:2009gi, vanBeest:2020kou, CarrenoBolla:2024fxy, Franco:2023flw, Arias-Tamargo:2024fjt, Franco:2023mkw, CarrenoBolla:2024fxy}, 
where white vertices on the boundary of the GTP indicate grouped $5$-branes ending on the same $7$-brane.
Let us denote the GTP by $\widetilde{\Delta}$, whereas $\Delta$ refers to an ordinary toric diagram of the same shape as $\widetilde{\Delta}$.

A central question is how the 
dynamics of the $5d$ $\mathcal N=1$ theory corresponding to a GTP $\widetilde{\Delta}$
is described by an integrable system.
We first consider the dual $(p,q)$-web corresponding to $\Delta$, 
where each external 5-brane ends on a distinct $7$-brane.
The multiple external $5$-branes
that are later identified with the same $7$-brane
initially correspond to distinct zig-zag paths in the brane tiling for $\Delta$.
These zig-zag paths parameterize the Casimirs and loop variables of the corresponding dimer integrable system.
Upon identifying the $5$-branes to the same $7$-brane, 
the corresponding zig-zag paths are set equal, 
which imposes constraints on the cluster variables of the 
original dimer integrable system.
This process is known as \textit{freezing},
and it effectively reduces the dimension of the phase space while fixing some of the 
Hamiltonians as constants.
In the $5d$ $\mathcal N=1$ theory, this corresponds to \textit{Higgsing}. 
The reduced system remains integrable: it inherits a Poisson structure 
and a set of commuting Hamiltonians from the parent dimer integrable system.

When a birational transformation $\varphi_A$ relates two toric diagrams $\Delta$ and $\Delta^\prime$
that have different numbers of internal points 
and hence correspond to dimer integrable systems with different numbers of Hamiltonians, 
we freeze canonical variables in the dimer integrable system for $\Delta^\prime$ 
so that it reduces to an integrable system corresponding to a GTP $\widetilde{\Delta}^\prime$ of the same shape as $\Delta^\prime$.
We show in an explicit example that the $(p,q)$-webs corresponding to $\Delta$ and $\widetilde{\Delta}^\prime$ are related by a Hanany-Witten transition
and that the dimer integrable system for $\Delta$ is birationally equivalent to the reduced integrable system corresponding to $\widetilde{\Delta}^\prime$.
In this letter, we demonstrate this mechanism explicitly for the birational transformation relating the $dP_1$ and $L^{2,5,1}$ models.

%========================================================
\section{The Integrable Systems for $dP_1$ and $L^{2,5,1}$}
\label{sec:03}

The $dP_1$ brane tiling and toric diagram $\Delta$ with one internal point are shown in the appendix.
The spectral curve of the associated dimer integrable system
takes the following form,
\beal{es03a01}
\Sigma ~:~
 \frac{z_1}{y}
+\frac{x}{y}
+ \frac{y}{z_2 z_3 x}
+\frac{y^2}{z_3 x}
-H
=0
~,~
\eea
where the 
monomial coefficients correspond to
Casimirs $\delta_{(n_x, n_y)}$ expressed in terms of the zig-zag paths $z_{1, \dots, 4}$.
The single internal point of the reflexive toric diagram for $dP_1$ corresponds to the 
Hamiltonian $H = \gamma_1 + \gamma_2 + \gamma_3 + \gamma_4$,
which is in terms of the following 1-loops,
\begin{align} \label{es03a02}
\gamma_1 &= \frac{z_1 z_2 f_3}{f_4}~,~ &
\gamma_2 &= z_1 z_2 f_3 ~,~ &
\nn\\
\gamma_3 &= z_1 z_2 f_2 f_3 ~,~ &
\gamma_4 &= \frac{z_1 z_2}{f_4} ~.~
\end{align}
The 1-loops $\gamma_{1,\dots,4}$ are expressed in terms of the zig-zag paths $z_{1, \dots, 4}$ and face loops $f_{1, \dots, 4}$,
whose expressions in terms of edge variables of the $dP_1$ brane tiling are given in
the appendix.

The $L^{2,5,1}$ brane tiling and toric diagram $\Delta^\prime$ with two internal points are summarized in the appendix.
The spectral curve of the dimer integrable system can be written in the following factorized form,
\beal{es03a10}
&&
\Sigma^\prime ~:~ 
\frac{1}{y} 
- H_1 
+ H_2 y
- \frac{y}{w_2 x}
\nn\\
&&
\hspace{1.5cm}
- (x+w_3)(x+w_4) \frac{w_1 y^2}{x}
= 0 
~,~
\eea
where the monomial coefficients correspond to the Casimirs $\delta^\prime_{(n_x,n_y)}$ 
written in terms of the zig-zag paths $w_{1,\dots,5}$ 
of the $L^{2,5,1}$ brane tiling, whose expressions in terms of edge variables are given in the appendix.
We note here that \eref{es03a10} coincides with the Seiberg-Witten curve of the associated $5d$ theory
$SU(3)_{2}+1F$ up to a standard reparameterization.
The two Hamiltonians corresponding to the two internal points of the 
$L^{2,5,1}$ toric diagram
are written in terms of 1-loops $\alpha_{1,2,3}$, $\beta_{1,2,3}$ and $\nu$
as follows,
\beal{es03a11}
H_1 &=&
\alpha_1 + \alpha_2 + \alpha_3
+ \beta_1 + \beta_2 + \beta_3 + \nu ~,~
\nn\\
H_2 &=&
\alpha_1 \alpha_2 + \alpha_2 \alpha_3 + \alpha_3 \alpha_1 
+ \alpha_1 \beta _3 + \alpha_2 \beta_1 + \alpha_3 \beta_2
\nn\\
&&
+ \beta_1 \beta_2 + \beta_2 \beta_3 + \beta_3 \beta_1
+ \alpha_2 \nu + \beta_2 \nu
~.~
\eea
The 1-loops can be expressed in terms of the zig-zag paths $w_{1,\dots,5}$
and face loops $h_{1, \dots, 7}$ of the $L^{2,5,1}$ brane tiling as shown in the appendix.
Here, we introduce a parameterization of face loops $h_{1, \dots, 6}$
in terms of a new auxiliary set of variables $\eta_{1, \dots, 6}$, as follows,
\beal{es03a12}
h_i = \left(\frac{w_4}{w_3}\right)^{\frac{(-1)^i}{3}}\frac{\eta_{i+2}}{\eta_i}
~,~
\eta_{i+6}=\eta_i 
~,~
\prod_{i=1}^6 \eta_i=1
~,~
\eea
with periodicity $\eta_{i+6} = \eta_i$.
Using this parameterization, we can write the 1-loops as follows, 
\beal{es03a15}
&
\alpha_m = \frac{(w_1 w_4)^{\frac{1}{3}}}{\eta_{2m} \eta_{2m+1}} 
~,~
\beta_m = \frac{(w_1 w_3)^{\frac{1}{3}}}{\eta_{2m-1} \eta_{2m}} 
~,~
\nu = \frac{(w_1^2 w_3 w_4)^{\frac{1}{3}} w_5}{(\eta_1)^2 \eta_2 \eta_6} 
~,~
&
\nn\\
\eea
where $m=1,2,3$. 

We can now introduce the following refined birational transformation, 
\beal{es03a20}
\varphi_A ~:~ (x,y) \mapsto (x, (x + z_1) y) ~,~
\eea
where $z_1$ corresponds to a zig-zag path in the $dP_1$ brane tiling. 
Under the above birational map, the spectral curve $\Sigma$ of the $dP_1$ model in \eref{es03a01}
becomes,
\beal{es03a21}
&&
\Sigma^\vee = \varphi_A \Sigma ~:~
\frac{1}{y} - H + \frac{1}{z_2 z_3} \left(x +  z_1\right) \frac{y}{x} 
\nn\\
&&
\hspace{2cm}
+ \frac{1}{z_3} \left( x + 2z_1 + \frac{z_1^2}{x}\right) y^2= 0 ~,~
\eea
whose Newton polygon $\Delta^\vee$ has the same shape as the toric diagram $\Delta^\prime$ for the $L^{2,5,1}$ model. 
The question we want to answer in this work is what the corresponding integrable system for $\Sigma^\vee$ is
and how it is related to the dimer integrable system with spectral curve $\Sigma^\prime$ in \eref{es03a10} corresponding to the $L^{2,5,1}$ model.

%========================================================
\section{The GTP Reduction and Birational Equivalence}
\label{sec:04}

We now show how the dimer integrable system for $L^{2,5,1}$ with spectral curve $\Sigma^\prime$ in \eref{es03a10}
can be reduced to the integrable system corresponding to $\Sigma^\vee$ in \eref{es03a21}.
To do so, we note that the $(p,q)$-web for the $L^{2,5,1}$ model
must have the $5$-branes corresponding to zig-zag paths $w_3$ and $w_4$ end on the same $7$-brane, 
imposing the constraint $w_3= w_4$. 
Under this identification, the resulting $(p,q)$-web is related to the $(p,q)$-web for the $dP_1$ model by a Hanany-Witten transition, as illustrated in \fref{fig_01}.

Under this constraint, the original Hamiltonians for the $L^{2,5,1}$ model in \eref{es03a11} can be expressed as, 
\begin{align}\label{es03a25}
H_1
\big|_{w_3=w_4}
=  &
~
\frac{(w_1 w_3)^{\frac{2}{3}} w_5}{(\eta_1)^2 \eta_2 \eta_6}
+ \sum_{m=1}^{6} \frac{(w_1 w_3)^{\frac{1}{3}}}{ \eta_m \eta_{m+1}}
~,~
\nn\\
H_2
\big|_{w_3=w_4}
= &
~
(w_1 w_3)^{\frac{2}{3}}
\Big(\sum_{i=1}^{3}\eta_{2i}\Big)
\Big(\sum_{j=1}^{3}\eta_{2j-1}\Big)
&
\nn\\
&
+ 
\frac{
w_1 w_3 w_5 (\eta_3+ \eta_5)
}{
\eta_1} 
~.~
\end{align}
To obtain an integrable system with the correct number of Hamiltonians, 
we impose the additional freezing constraint
$\eta_1+\eta_3+\eta_5=0$, which yields,
\begin{align} \label{es03a26}
\widetilde{H}_1 
=&~
\frac{(w_1 w_3)^{\frac{2}{3}}w_5}{\eta_1^2 \eta_2 \eta_6}
\nn\\
&
-(w_1 w_3)^{\frac{1}{3}}
\left(
\frac{\eta_5}{\eta_1 \eta_2 \eta_3}
+\frac{\eta_1}{\eta_3 \eta_4 \eta_5}
+\frac{\eta_3}{\eta_5 \eta_6 \eta_1}
\right)
~,~
\nn\\
\widetilde{H}_2 
=&
-w_1 w_3 w_5
~,~
\end{align}
where we can see that $\widetilde{H}_2$ is now a constant.
leaving only a single dynamical Hamiltonian $\widetilde{H}_1$ after freezing.

The resulting spectral curve of the reduced integrable system corresponds to the GTP 
with the same shape as the toric diagram for $L^{2,5,1}$.
It takes the following form, 
\beal{es03a28}
\widetilde{\Sigma}^\prime ~:~
&
\frac{1}{y}
- \widetilde{H}_1
- w_1 w_3 w_5 (x + w_3) \frac{y}{x}
\nn\\
&
- w_1 \left(x  + 2 w_3 + \frac{w_3^2}{x} \right) y^2 = 0
~,~
\eea
By comparing the spectral curve $\widetilde{\Sigma}^\prime$ above with $\Sigma^\vee$ in \eref{es03a21}
obtained from birationally transforming the spectral curve $\Sigma$ for the $dP_1$ model, 
we obtain the following identifications between zig-zag variables,
\beal{es03a30}
z_1=w_3 ~,~
z_2=\frac{1}{w_3 w_5}~,~
z_3=-\frac{1}{w_1}~,~
z_4= -w_1 w_5~,~
\nn\\
\eea
together with the map between the 1-loops $\gamma_{1,\dots,4}$ from the $dP_1$ model 
and the 1-loops in the reduced $L^{2,5,1}$ model,
expressed in terms of $w_{1,\dots,5}$ and $\eta_{1,\dots,6}$, as follows,
\begin{align}\label{es03a31}
\gamma_1 &= 
\frac{( w_1 w_3)^{\frac{2}{3}} w_5}{ (\eta_1)^2 \eta_2 \eta_6}
~,~
&
\gamma_2 &= 
- \frac{(w_1 w_3)^{\frac{1}{3}} \eta_3}{ \eta_5 \eta_6 \eta_1} 
~,~
\nn\\
\gamma_3 &= 
- \frac{(w_1 w_3)^{\frac{1}{3}} \eta_1}{\eta_3 \eta_4 \eta_5}
~,~&
\gamma_4 &= 
- \frac{(w_1 w_3)^{\frac{1}{3}} \eta_5}{\eta_1 \eta_2 \eta_3}
~.~
\end{align}
Using $\prod_{m=1}^6\eta_m=1$ and \eref{es03a12}, the 1-loops $\gamma_{1,\dots,4}$ above can be expressed as products of
the original face variables $h_i$ of the $L^{2,5,1}$ model.

We note here that the Poisson brackets between $\gamma_{1, \dots, 4}$ in \eref{es03a31}
satisfy the original Poisson brackets from the $dP_1$ model.
This implies that the dimer integrable system for the $dP_1$ model with spectral curve $\Sigma$ in \eref{es03a01}
is birationally equivalent to the reduced integrable system with spectral curve $\widetilde{\Sigma}^\prime$ in \eref{es03a28},
obtained by freezing the original dimer integrable system for the $L^{2,5,1}$ model with spectral curve $\Sigma^\prime$ in \eref{es03a10}. 

When we look at the exponents of $x,y \in \mathbb{C}^*$ in the terms in $\widetilde{\Sigma}^\prime$ in \eref{es03a28}
we see that they correspond to vertices of the GTP with the same shape as the toric diagram for the $L^{2,5,1}$ model.
Considering the associated $5d$ $\mathcal{N}=1$ pure $SU(2)_{3\pi}$ theory,
its Coulomb branch is described by a non-convex polygon embedded inside the GTP, 
such that one of the internal vertices of the original toric diagram for $L^{2,5,1}$ corresponding to $\widetilde{H}_2$ becomes a boundary vertex in the non-convex polygon,
while the only internal vertex of the latter corresponds to the single dynamical Hamiltonian $\widetilde{H}_1$ of the reduced integrable system.
As such, we can see that the reduced integrable system correctly describes the dynamics of the $5d$ $\mathcal{N}=1$ pure $SU(2)_{3\pi}$ theory corresponding to the GTP and the embedded non-convex polygon shown in \fref{fig_01}.

From the viewpoint of the integrable system, our work provides a concrete construction of an integrable system associated with a GTP, 
and equivalently with an embedded non-convex polygon that describes the $5d$ $\mathcal{N}=1$ theory with Coulomb moduli.
Our work thus extends the construction of these integrable systems beyond the original scope of dimer integrable systems first introduced in \cite{2011arXiv1107.5588G, Eager:2011dp}.
Moreover, from the viewpoint of birational transformations, our example shows
that when the birational map changes the number of internal points from one in the $dP_1$ model to two in the $L^{2,5,1}$ model, 
the reduced integrable system corresponding to the spectral curve $\widetilde{\Sigma}^\prime$ in \eref{es03a28}
and the GTP is nevertheless birationally equivalent to the original dimer integrable system corresponding to the $dP_1$ model.
Our work therefore shows that birational transformations that change the number of internal points in toric diagrams
still define birational equivalences between corresponding integrable systems. 

%========================================================
\section{Summary and outlook}

In our work, 
we have identified a natural extension of birational equivalence between dimer integrable systems 
to the case in which the birational map
changes the number of internal points of the toric diagram.
The key observation is that the dual $(p,q)$-web
admits brane configurations in which multiple external 5-brane legs end on a common $7$-brane after a Hanany-Witten move.
This leads to GTPs whose additional boundary information allows us to introduce freezing of the integrable system corresponding to the original toric diagram.
The resulting reduced integrable system is birationally equivalent to the original dimer integrable system prior to the birational transformation.
Moreover, the reduced integrable system corresponds naturally to the GTP and describes the dynamics of the Higgsed $5d$ $\mathcal{N}=1$ theory.
We expect our observations to extend beyond the example based on the $dP_1$ and $L^{2,5,1}$ models.
In future work, 
we plan to construct new integrable systems corresponding to a variety of GTPs and non-convex polygons, 
while reporting on a new family of birational equivalences.

\paragraph{Acknowledgements.}
The authors would like to thank 
for hospitality by the Simons Center for Geometry and Physics during the Simons Physics Workshops,
where this project was initiated and conducted. 
K. L. is supported in part by the BIMSA start-up fund and the Beijing Natural Science Foundation International Scientist Project No. IS25024.
N. L. is supported by the Institute of Basic Science (IBS) under Project No. IBS-R003-D1.
R.-K. S. is supported by an Outstanding Young Scientist Grant (RS-2025-00516583) of the National Research Foundation of Korea (NRF). He is also partly supported by the BK21 Program (“Next Generation Education Program for Mathematical Sciences”, 4299990414089) funded by the Ministry of Education in Korea and the National Research Foundation of Korea (NRF).

%%%%%%%%%%%%%%%%%%%%%%%%%%%%%%%%%%%%%%%%%%%%%%%%%%%%%%%%%%%%%%%%%
%%%
%%%                     BIBLIOGRAPHY
%%%
%%%%%%%%%%%%%%%%%%%%%%%%%%%%%%%%%%%%%%%%%%%%%%%%%%%%%%%%%%%%%%%%%

% \bibliographystyle{jhep}
\bibliography{mybib}

%========================================================
\onecolumngrid
\section*{Supplementary Materials}
\twocolumngrid

%--------------------------------------------------------------------------
\subsection{The $dP_1$ Model}
\label{app:01}

%------------------------------------------------------------------------------------------
\begin{figure}[H]
\centering
\includegraphics[width=0.8\linewidth]{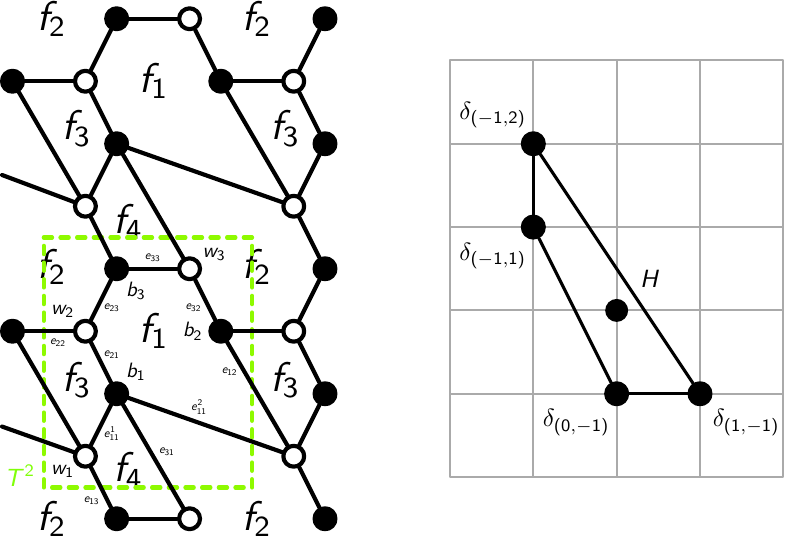}
\caption{The brane tiling and toric diagram for $dP_1$.}
\label{fig_02}
\end{figure}
%------------------------------------------------------------------------------------------

The brane tiling and the toric diagram $\Delta$ for the $\text{dP}_1$ model \cite{Feng:2002zw, Feng:2001xr, Benvenuti:2006qr, Hanany:2012hi} are shown in \fref{fig_02}.
For the choice of fundamental domain on $T^2$ shown in \fref{fig_02}, 
the Kasteleyn matrix, which encodes the adjacency structure of the bipartite graph of the brane tiling,
takes the form,
\beal{es10a01}
K = 
\left(
\begin{array}{ccc}
- e_{11}^{1} + \frac{e_{11}^{2}}{x} & \frac{e_{12}}{x} & \frac{e_{13}}{y} \\
e_{21} & \frac{e_{22}}{x} & e_{23} \\
e_{31}{y} & e_{32}  & e_{33}
\end{array}
\right)
~,~
\eea
where $e_{ij}$ are edge variables assigned to 
edges connecting white nodes $w_i$
and black nodes $b_j$ in the brane tiling.
Taking the determinant of $K_{\text{dP}_1}$,
acting with an $SL(2,\BZ)$ transformation on the coordinates $x,y \in \mathbb{C}^*$
and followed by a shift of the origin,
we obtain the Newton polynomial in the form, 
\beal{es10a02}
&&
P(x,y) 
= \bar{p}_0 \Big[ \d_{(0,-1)}\frac{1}{y} + \d_{(1,-1)} \frac{x}{y} 
~,~
\nn\\
&&
\hspace{2cm}
+ \d_{(-1,1)} \frac{y}{x} + \d_{(-1,2)} \frac{y^2}{x} - H \Big]
~,~
\eea
where the reference perfect matching $\bar{p}_0 = e_{11}^{2+}e_{22}^+e_{33}^+$ has been factored out. 

The 1-loops and Casimirs can be expressed in terms of zig-zag paths given by,
\beal{es10a04}
z_1 &=& (e^+_{12}~e^-_{22}~e^+_{23}~e^-_{33}~e^+_{31}~e_{11}^{2-})
~,~\nn\\
z_2 &=& (e^+_{13}~e^-_{23}~e^+_{21}~e_{11}^{1-})
~,~\nn\\
z_3 &=& (e^{2+}_{11}~e^-_{21}~e^+_{22}~e^-_{32}~e^+_{33}~e^-_{13})
~,~\nn\\
z_4 &=& (e_{11}^{1+}~e^-_{31}~e^+_{32}~e^-_{12})
~,~
\eea
and face loops given by,
\beal{es10a05}
f_1 &=& (e_{33}^+~e_{23}^-~e_{21}^+~e_{11}^{2-}~e_{12}^+~e_{32}^-)
~,~ \nn\\
f_2 &=& (e_{11}^{2+}~e_{31}^-~e_{32}^+~e_{22}^-~e_{23}^+~e_{13}^-)
~,~ \nn\\
f_3 &=& (e_{22}^+~ e_{12}^-~e_{11}^{1+}~e_{21}^-)
~,~ \nn\\
f_4 &=& (e_{31}^+~e_{11}^{1-}~e_{13}^+~e_{33}^-)~.~
\eea
The face loops and zig-zag paths are not independent and satisfy the following relations, 
\beal{es10a06}
z_1 z_2 z_3 z_4 = 1~,~  
z_1 z_3^2 z_4^3 = \frac{f_2 f_3^2}{f_4} ~,~
z_1^2 z_2^3 z_3= \frac{f_1 f_4^2}{f_3}
~.~
\eea

The Casimirs, corresponding to monomial coefficients in \eref{es10a02}, 
can be expressed in terms of zig-zag paths as follows, 
\begin{align}\label{es10a07}
\delta_{(0,-1)} &= z_1 ~,~ & 
\delta_{(1,-1)} &= 1~,~ &
\nn\\
\delta_{(-1,1)} &= \frac{1}{z_2z_3}~,~ &
\delta_{(-1,2)} &= \frac{1}{z_3}~.~
\end{align}
The 1-loops in the Hamiltonian $H = \gamma_1 + \gamma_2 + \gamma_3 + \gamma_4$
can be expressed in terms of zig-zag paths and face loops as given in \eref{es03a02}.
The 1-loops satisfy the following commutation relations,
\begin{align}\label{es10a10}
\{\gamma_i,\gamma_{i+1}\} &= \gamma_i\gamma_{i+1}~,~ &
\{\gamma_i,\gamma_{i+3}\} &= -\gamma_i\gamma_{i+3}~,~ 
\nn \\
\{\gamma_2,\gamma_4\} &= -\gamma_2\gamma_4~,~ &
\{\gamma_1,\gamma_3\} &= 0~,~
\end{align}
where $i=1, \dots 4$ and, by periodicity, $\gamma_i = \gamma_{i+4}$.
We also note that the two following combinations of 1-loops 
lie in the center of the Poisson algebra,
\beal{es10a11}
\gamma_1 \gamma_3 = z_1 z_4~,~ 
\gamma_2 \gamma_3 \gamma_4 = \frac{z_1}{z_3}
~.~
\eea

%--------------------------------------------------------------------------
\subsection{The $L^{2,5,1}$ Model}
\label{app:02}

%------------------------------------------------------------------------------------------
\begin{figure}[H]
\centering
\includegraphics[width=0.8\linewidth]{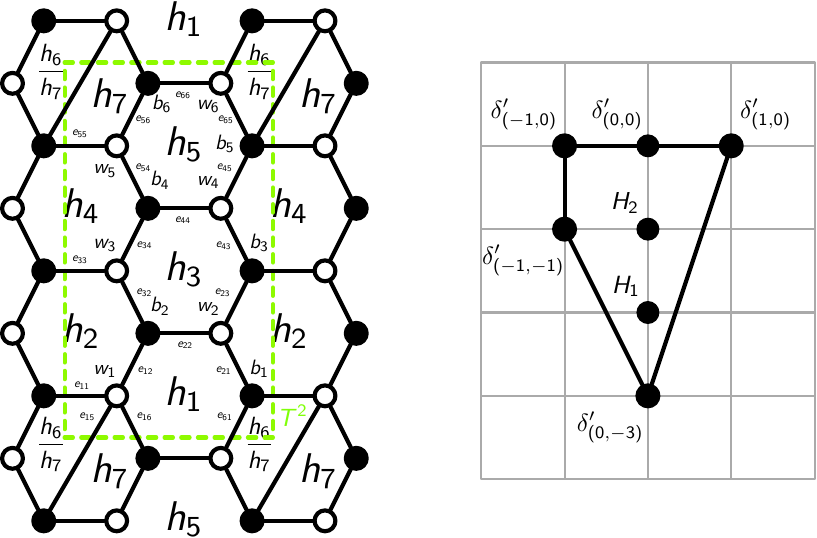}
\caption{The brane tiling and toric diagram for $L^{2,5,1}$.}
\label{fig_03}
\end{figure}
%------------------------------------------------------------------------------------------

The brane tiling and toric diagram $\Delta^\prime$ for the $L^{2,5,1}$ model \cite{Cvetic:2005ft, Franco:2005sm, Benvenuti:2005ja, Butti:2005sw}
are shown in \fref{fig_03}.
The corresponding Kasteleyn matrix
is as follows, 
\beal{es20a01}
K^\prime
= 
\left(
\begin{array}{cccccc}
\frac{e_{11}}{x} & e_{12} & 0 & 0 & -\frac{e_{15}}{xy} & \frac{e_{16}}{y} \\
e_{21} & e_{22} & e_{23} & 0 & 0 & 0 \\
0 & e_{32} & \frac{e_{33}}{x} & e_{34} & 0 & 0 \\
0 & 0 & e_{43} & e_{44} & e_{45} & 0 \\
0 & 0 & 0 & e_{54} & \frac{e_{55}}{x} & e_{56} \\
{e_{61}}{y} & 0 & 0 & 0 & e_{65} & e_{66} 
\end{array}
\right)
~.~
\eea
Under a $GL(2,\mathbb{Z})$ transformation and a shift of the origin, the Newton polynomial takes the following form, 
\beal{es20a02}
&&
P^\prime(x,y) 
= \bar{p}_0 \Big [ 
\delta_{(0,-1)} \frac{1}{y} 
- H_1 
+ H_2 y
- \delta_{(-1,1)} \frac{y}{x}
\nn\\
&&
\hspace{2.0cm}
- \delta_{(1,2)} x y^2
- \delta_{(0,2)} y^2 
- \delta_{(-1,2)} \frac{y^2}{x} \Big]
~,~
\eea
where the factored-out reference perfect matching weight is given by $\bar{p}_0 = e_{11}^+e_{22}^+e_{33}^+e_{44}^+e_{55}^+e_{66}^+$.
The 1-loops and Casimirs can be expressed in terms of zig-zag paths of the form,
\beal{es20a03}
w_1 &=& (e_{12}^+~e_{22}^-~e_{23}^+~e_{33}^-~e_{34}^+~e_{44}^-~e_{45}^+~e_{55}^-~e_{56}^+~e_{66}^-~e_{61}^+~e_{11}^-)~,~\nn\\
w_2 &=& (e_{11}^+~e_{21}^-~e_{22}^+~e_{32}^-~e_{33}^+~e_{43}^-~e_{44}^+~e_{54}^-~e_{55}^+~e_{15}^-)~,~\nn\\
w_3 &=& (e_{16}^+~e_{56}^-~e_{54}^+~e_{34}^-~e_{32}^+~e_{12}^-)~,~\nn\\
w_4 &=& (e_{21}^+~e_{61}^-~e_{65}^+~e_{45}^-~e_{43}^+~e_{23}^-)~,~\nn\\
w_5 &=& (e_{15}^+~e_{65}^-~e_{66}^+~e_{16}^-)~,~
\eea
and face loops given by,
\beal{es20a04}
h_i &=& (e_{[i+1],[i+1]}^+ ~ e_{i,[i+1]}^- ~ e_{i,[i-1]}^+
~ e_{[i-1],[i-1]}^- 
\nn\\
&&
\hspace{0.5cm}
~ e_{[i-1],i}^+ ~ e^-_{[i+1],i})
~,~\nn\\
h_7 &=&(e^+_{15}~e_{55}^- ~e_{56}^+ ~e_{16}^-)
~,~
\eea
where $i=1, \dots 6$.
Here, for convenience, the faces in the brane tiling in \fref{fig_03}
are labelled by $h_1, \dots, h_5$, $h_7$ and $h_6/h_7$, 
where $h_6$ defines a closed directed path in the brane tiling enclosing two faces corresponding to $h_7$ and $h_6/h_7$. 
Using this choice of parameterization, 
the face variables satisfy the following Poisson commutation relations, 
\beal{es20a05}
\{h_i,h_j\} &=& (2\delta_{i,j+1}-2\delta_{i,j-1}+\delta_{i,j-2}-\delta_{i,j+2}) h_i h_j 
~,~ \nn\\
\{h_i, h_7\} &=& (\delta_{i,1}-\delta_{i,6} - \delta_{i,5}+\delta_{i,4})h_i h_7
~.~
\eea
The face loops and zig-zag paths form the following relations, 
\beal{es20a06}
&
w_1 w_2 w_3 w_4 w_5=1 ~,~
\frac{h_2 h_6^2}{h_3 h_5^{2} h_7^{3}}=w_1^2 w_2^3 w_3 w_4^4 ~,~
&
\nn\\
&
h_1 h_3 h_5 = \frac{w_3}{w_4} ~,~
h_2 h_4 h_6 = \frac{w_4}{w_3} ~,~
&
\nn\\
&
\frac{h_1}{h_2 h_4^{2} h_5 h_7^{3}} =w_1^2 w_2^3 w_3^4 w_4
~.~
\eea

In \eref{es03a12}, we introduced a reparameterization of the face loops $h_{1, \dots, 6}$
in terms of auxiliary variables $\eta_{1,\dots, 6}$, which need to satisfy the 
commutation relations between the face loops. 
We first write, 
\beal{es20a20}
\{\eta_i,\eta_j\} = \epsilon_{ij} \eta_i \eta_j
~,~
\eea
where in order for the above commutation relations to satisfy the
commutation relations between the original face loops as well as the constraint $\prod_{i=1}^6 \eta_i = 1$, 
we have,
\beal{es20a21}
&
\epsilon_{14}=1+\epsilon_{12} - \epsilon_{24} ~,~
\epsilon_{16} = 1+\epsilon_{12}-\epsilon_{26}~,~ 
&
\nn\\
&
\epsilon_{23} = -1-\epsilon_{12}-\epsilon_{13}~,~ 
\epsilon_{34}=\epsilon_{12}+\epsilon_{13}-\epsilon_{24}~,~ 
&
\nn\\
&
\epsilon_{36}=\epsilon_{12} + \epsilon_{13}- \epsilon_{26}~,~
\epsilon_{45}=2\epsilon_{12}+\epsilon_{13}-\epsilon_{26}~,~ 
&
\nn\\
&
\epsilon_{46} = 1+\epsilon_{26}-\epsilon_{24}~,~ 
\epsilon_{56}=-2-2\epsilon_{12} - \epsilon_{13}+\epsilon_{24}~,~ 
&
\nn\\
&
\epsilon_{15}=-2-3\epsilon_{12} - \epsilon_{13}+\epsilon_{24} + \epsilon_{26}~,~ 
&
\nn\\
&
\epsilon_{25}=1+2\epsilon_{12} +\epsilon_{13}-\epsilon_{24} - \epsilon_{26}~,~ 
&
\nn\\
&
\epsilon_{35}=-1-3\epsilon_{12} - 2\epsilon_{13}+\epsilon_{24} + \epsilon_{26}~,~
&
\eea
with four free parameters $\epsilon_{12}$, $\epsilon_{13}$, $\epsilon_{24}$, $\epsilon_{26}$.
By making an appropriate choice for these parameters, we obtain the following 
commutation relations,
\beal{es20a22}
\frac{\{\eta_i,\eta_j\}}{\eta_i \eta_j} = 
\frac{2}{3}(\d_{i,j+1}-\d_{i,j-1}) + \frac{1}{3} (\d_{i,j-2}-\d_{i,j+2})
~.~
\nn\\
\eea

\end{document}